\documentclass[prl,twocolumn,english]{revtex4-1}
\usepackage[T1]{fontenc}
\usepackage{color}
\usepackage{amsmath}
\usepackage{amssymb}
\usepackage{graphicx}
\usepackage{esint}
\usepackage{lineno}

\definecolor{ablue}{rgb}{0.1,0.35,0.75}
\definecolor{agreen}{rgb}{0,0.6,0.4}

\usepackage[normalem]{ulem}

\makeatletter
\@ifundefined{textcolor}{}
{%
 \definecolor{BLACK}{gray}{0}
 \definecolor{WHITE}{gray}{1}
 \definecolor{RED}{rgb}{1,0,0}
 \definecolor{GREEN}{rgb}{0,1,0}
 \definecolor{BLUE}{rgb}{0,0,1}
 \definecolor{CYAN}{cmyk}{1,0,0,0}
 \definecolor{MAGENTA}{cmyk}{0,1,0,0}
 \definecolor{YELLOW}{cmyk}{0,0,1,0}
}

\renewcommand{\fnum@figure}{\textbf{Figure~\thefigure}}
\usepackage{units}
\usepackage{lipsum}
\usepackage{babel}

\newcommand{\vecPar}[1]{{\mathbf{#1}_\parallel}}

\makeatother

\usepackage{babel}
\begin{document}

\title{Phonon-assisted Photoluminescence from Dark Excitons in Monolayers of Transition Metal Dichalcogenides}

\author{Samuel Brem$^1$}
\author{August Ekman$^1$}
\author{Dominik Christiansen$^2$}
\author{Florian Katsch$^2$}
\author{Malte Selig$^2$}
\author{Cedric Robert$^3$}
\author{Xavier Marie$^3$}
\author{Bernhard Urbaszek$^3$}
\author{Andreas Knorr$^2$}
\author{Ermin Malic$^1$}
\affiliation{$^1$Chalmers University of Technology, Department of Physics, 41296 Gothenburg, Sweden}
\affiliation{$^2$Technical University Berlin, Institute of Theoretical Physics, 10623 Berlin, Germany}
\affiliation{$^3$Universit\'e de Toulouse, INSA-CNRS-UPS, LPCNO, 31077 Toulouse, France}

\begin{abstract}
The photoluminescence (PL) spectrum of transition metal dichalcogenides (TMDs) shows a multitude of emission peaks below the bright exciton line and not all of them have been explained yet. Here, we study the emission traces of phonon-assisted recombinations of momentum-dark excitons. To this end, we develop a microscopic theory describing simultaneous exciton, phonon and photon interaction and including consistent many-particle dephasing. We explain the drastically different PL below the bright exciton in tungsten- and molybdenum-based materials as result of different configurations of bright and dark states. In good agreement with experiments, we show that WSe$_2$ exhibits clearly visible low-temperature PL signals stemming from the phonon-assisted recombination of momentum-dark excitons. 
\end{abstract}
\maketitle

\begin{figure}[t!]
\includegraphics[width=80mm]{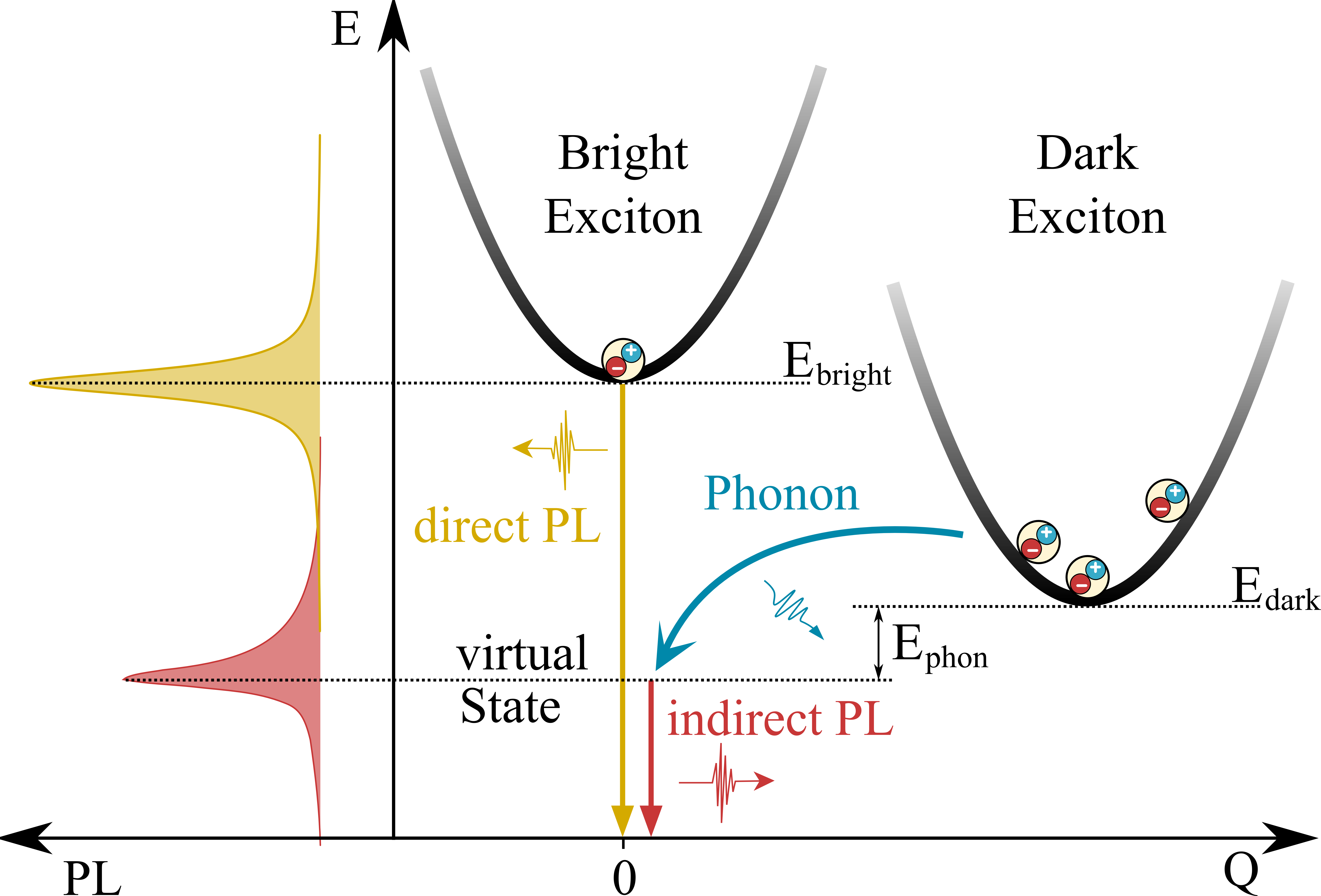}
\caption{Sketch of direct and indirect decay channels for excitons showing the underlying scattering processes in the excitonic center-of-mass dispersion (right) and the corresponding PL signals (left). Momentum-dark excitons can decay by emitting or absorbing a phonon (blue arrow) and subsequently emitting a photon, which contributes to the indirect PL signal (red arrow).}
\label{fig:scheme} 
\end{figure}

The cryogenic photoluminescence (PL) spectrum of two dimensional semiconductors provides a powerful tool to study intriguing quantum phenomena invisible at room temperature or in linear optical experiments. The large variety of low temperature emission features at energies below the bright exciton resonance indicates the existence of bound exciton configurations, such as trions, biexcitons \cite{mak2013tightly, wang2014valley, courtade2017charged,  ye2018efficient, barbone2018charge} and trapped excitons \cite{tonndorf2015single, zhang2017defect}, but could also result from the indirect recombination of intrinsically dark exciton states \cite{ lindlau2017identifying, koperski2017optical, mueller18, malic18, selig2016excitonic, feierabend17}. In particular, the radiative decay of a momentum indirect electron-hole pair requires the simultaneous interaction with a phonon to fulfill the momentum conservation and is therefore very inefficient compared to the direct recombination of an exciton with zero center-of-mass momentum. However, in indirect semiconductors, where the momentum indirect exciton is located below the bright state, the low temperature emission can exhibit strong phonon-assisted signals due to a large population of dark states, as depicted in Fig. \ref{fig:scheme}. Recently, the PL spectrum of hexagonal boron nitride has been shown to be dominated by phonon assisted processes \cite{cassabois2016hexagonal} resulting from an indirect bandgap. Several theoretical and experimental studies have demonstrated that in tungsten-based monolayer materials intervalley excitons are located below the optically bright exciton \cite{zhang2015experimental,echeverry2016splitting, selig2016excitonic,berghaeuser17} and recent experimental PL studies on hBN-encapsulated tungsten diselenide have revealed a multitude of low-temperature emission peaks whose microscopic origin still needs to be clarified \cite{ courtade2017charged,lindlau2017identifying, ye2018efficient,barbone2018charge}.

While the radiative decay of bright excitons has been extensively studied \cite{singh2015intrinsic, palummo2015exciton, steinhoff2015efficient, pollmann2015resonant, robert2016exciton, wang2016radiative}, the theoretical description of simultaneous exciton, phonon and photon interactions remains challenging because of the non-markovian nature of these processes. In previous studies, specific cases, such as optical phonon replicas of bright excitons, have been theoretically investigated within the polaron-frame \cite{feldtmann2009phonon, feldtmann2010theoretical} and non-markovian treatments of the full density matrix \cite{chernikov2012phonon,shree2018observation}. However, there is so far no general theoretical framework for the phonon-assisted exciton recombination including both optical and acoustic phonons as well as intra- and intervalley recombination channels on a microscopic footing. Moreover, the underlying dephasing rates shaping the phonon-assisted emission signals have not been derived in literature yet. Here, we present a fully microscopic model of the phonon-assisted PL from dark excitons. Based on the fundamental equations-of-motion of the many-particle density matrix we find an analytical formula which allows to calculate temperature-dependent photoluminescence spectra for thermalized exciton gases. We apply the derived formula to calculate the luminescence spectrum of hBN-encapsulated TMD monolayers. For tungsten diselenide (WSe$_2$) we are able to explain so far observed but unidentified low temperature emission features between $50$ and $\unit[80]{meV}$ below the bright exciton \cite{courtade2017charged,lindlau2017identifying,heinz18}, by assigning them to the phonon assisted recombination of momentum-dark excitons.  In contrast, for molybdenum diselenide (MoSe$_2$) we find no additional peaks and an opposite asymmetric broadening of the bright exciton resonance compared to WSe$_2$, which can be explained by the absence of lower lying dark excitons. Overall, our work provides new insights into phonon-assisted exciton luminescence and can be applied to determine emission spectra of arbitrary semiconducting materials.\\

\textit{Theoretical approach:} The PL signal can be derived from the many-particle density matrix of an interacting system of electrons, phonons and photons. Throughout this work we assume a low excitation density, where the fermionic substructure of excitons can be neglected, and excitons are well described as a gas of non-interacting bosons. In this regime the many-particle physics of an undoped monolayer can be described by the excitonic Hamiltonian \cite{toyozawa1958theory, ivanov1993self, chernikov2012phonon, katsch2018theory}, which in rotating frame reads
\begin{eqnarray} \label{eq:Hamilton}
H=\sum &\bigg{(}& E^{\mu}_{\vecPar{Q}} a^{\dagger}_{\mu \vecPar{Q}}a^{\phantom\dagger}_{\mu \vecPar{Q}} + \Omega^{\alpha}_{\vecPar{q}} b^{\dagger}_{\alpha \vecPar{q}}b^{\phantom\dagger}_{\alpha \vecPar{q}} + \omega^{\sigma}_{\mathbf{k}} c^{\dagger}_{\sigma \mathbf{k}}c^{\phantom\dagger}_{\sigma \mathbf{k}} \nonumber\\ 
  &+& M^{\mu}_{\sigma \mathbf{k}} c^{\dagger}_{\sigma \mathbf{k}}a^{\phantom\dagger}_{\mu \vecPar{k}}+M^{\mu\ast}_{\sigma \mathbf{k}} a^{\dagger}_{\mu \vecPar{k}} c^{\phantom\dagger}_{\sigma \mathbf{k}} \nonumber \\
  &+& D^{\nu\mu}_{\alpha \vecPar{q}} a^{\dagger}_{\nu \vecPar{Q}+\vecPar{q}}a^{\phantom\dagger}_{\mu \vecPar{Q}}(b^{\phantom\dagger}_{\alpha \vecPar{q}}+b^{\dagger}_{\alpha, -\vecPar{q}})  \bigg{).}
\end{eqnarray}
Here, $\vecPar{k}$ denotes the component of the vector $\mathbf{k}$ parallel to the monolayer plane and the summation is performed over all appearing quantum numbers. We use annihilation (creation) operators $a^{(\dagger)}_{\mu \vecPar{Q}}$, $b^{(\dagger)}_{\alpha \vecPar{q}}$ and $c^{(\dagger)}_{\sigma \mathbf{k}}$ for excitons in the state $\mu$, phonons in the mode $\alpha$, and photons with the polarization $\sigma$, respectively, while the vectorial quantum numbers denote the momentum of the particle. Moreover, the corresponding dispersions for the three particle species are given by  $E^{\mu}_{\vecPar{Q}}$, $\Omega^{\alpha}_{\vecPar{q}}$ and $\omega^{\sigma}_{\mathbf{k}}$. Here, the excitonic bandstructure $E^{\mu}_{\vecPar{Q}}$ is decomposed into valleys around energetic extrema, where excitons can be described as free particles with effective valley masses. The exciton index $\mu$ acts as a compound index containing main, angular, spin and valley quantum numbers. 
The second line of Eq. \ref{eq:Hamilton} describes the conversion of excitons to photons and vice versa under conservation of the in-plane momentum, whose probability is determined by the exciton-photon matrix element $M^{\mu}_{\sigma \mathbf{k}}$. Finally, excitons can scatter from the state $(\mu, \vecPar{Q})$ to $(\nu, \vecPar{Q}+\vecPar{q})$ by emitting or absorbing a phonon, guided by the exciton-phonon matrix element $D^{\nu\mu}_{\alpha \vecPar{q}}$ \cite{selig2016excitonic, selig2018dark, brem2018exciton}. For convenience, we will  drop the index $\parallel$ in the following keeping in mind that exciton and phonon momenta are two dimensional. Moreover, we suppress the photon and phonon mode index for a better readability. 

In this framework, the PL can be determined from the temporal change of the photon density $n_{\mathbf{k}}=\langle  c^{\dagger}_{\mathbf{k}}c^{\phantom\dagger}_{\mathbf{k}}\rangle$ \cite{kira1999quantum}. Applying the Heisenberg equation of motion we find coupled differential equations for the photon density, the polarization $\mathcal{S}^\mu_\mathbf{k}=\langle c^{\dagger}_{\mathbf{k}}a^{\phantom\dagger}_{\mu \vecPar{k}}\rangle$, the phonon-assisted polarization $\mathcal{U}^{\mu,\pm}_{\mathbf{kq}}=\langle c^{\dagger}_{\mathbf{k}}b^{(\dagger)}_{\mp\mathbf{q}}a^{\phantom\dagger}_{\mu \vecPar{k}-\mathbf{q}}\rangle$ and the exciton-phonon correlation $\mathcal{C}^{\nu\mu,\pm}_{\mathbf{kq}}=\langle a^{\dagger}_{\nu \vecPar{k}} a^{\phantom\dagger}_{\mu \vecPar{k}-\mathbf{q}}b^{(\dagger)}_{\mp\mathbf{q}}\rangle$: 
\begin{eqnarray} 
\dfrac{d}{dt} n_{\mathbf{k}} &=& \dfrac{2}{\hbar}\sum_\mu\Im\text{m}\{M^{\mu}_{\mathbf{k}}\mathcal{S}^\mu_\mathbf{k}\}, \label{eq:EoM_n} \\
i \hbar \dfrac{d}{dt}\mathcal{S}^\mu_\mathbf{k} &=& (E^{\mu}_{\vecPar{k}}-\omega_{\mathbf{k}})\mathcal{S}^\mu_\mathbf{k} - M^{\mu\ast}_{\mathbf{k}}N^\mu_{\vecPar{k}} \nonumber \\
           &+& \sum_{\nu\mathbf{q}\pm} D^{\mu\nu}_{\mathbf{q} }\mathcal{U}^{\mu\pm}_{\mathbf{kq}} + \dot{\mathcal{S}}\rvert^\text{rad}_\text{deph.}, \\
i \hbar \dfrac{d}{dt}  \mathcal{U}^{\nu,\pm}_{\mathbf{kq}} &=& (E^\nu_{\vecPar{k}-\mathbf{q}}\mp \Omega_\mathbf{q} -\omega_\mathbf{k}) \mathcal{U}^{\nu,\pm}_{\mathbf{kq}} -\sum_\mu M^{\mu\ast}_{\mathbf{k}} \mathcal{C}^{\mu\nu,\pm}_{\mathbf{kq}} \nonumber \\
           &+& \sum_\mu D^{\mu\nu\ast}_{\mathbf{q}} \eta^{\pm}_\mathbf{q} \mathcal{S}^\mu_\mathbf{k} + \dot{\mathcal{U}}\rvert^\text{phon}_\text{deph.}, \label{eq:Udot}\\
i \hbar \dfrac{d}{dt} \mathcal{C}^{\nu\mu,\pm}_{\mathbf{kq}} &=& (E^\mu_{\vecPar{k}-\mathbf{q}}-E^\nu_{\vecPar{k}}\mp \Omega_\mathbf{q}) \mathcal{C}^{\nu\mu,\pm}_{\mathbf{kq}} \nonumber \\
          &-& D^{\nu\mu\ast}_{\mathbf{q}}Q^{\nu\mu,\pm}_{\mathbf{kq}} + \dot{\mathcal{C}}\rvert^\text{rad+phon}_\text{deph.}, \label{eq:Cdot}
\end{eqnarray}
where $\eta^{\pm}_\mathbf{q}=1/2\mp 1/2 +\langle b^{\dagger}_{\mathbf{q}} b^{\phantom\dagger}_{\mathbf{q}}\rangle$ denotes the relevant phonon occupation factor for absorption/emission, $N^\mu_{\mathbf{Q}}=\langle a^{\dagger}_{\mu \mathbf{Q}}a^{\phantom\dagger}_{\mu \mathbf{Q}}\rangle$ represents the exciton occupation and $Q^{\nu\mu,\pm}_{\mathbf{kq}}=\eta^{\mp}_\mathbf{q} N^\mu_{\vecPar{k}-\mathbf{q}}-\eta^{\pm}_\mathbf{q} N^\nu_{\vecPar{k}}$ is the source of the corresponding exciton-phonon correlation.

To obtain the above set of equations we have factorized appearing many-particle expectation values according to the cluster expansion scheme \cite{kira2006many}. Here, we neglect coherent quantities and non-linear density dependencies, assuming a fast decoherence after the initial laser excitation and low exciton densities. Moreover, we disregard contributions connected to multi-phonon processes, cf. supplementary. Important higher order correlations have been abbreviated with $(\cdot)\rvert^\text{phon}_\text{deph.}$. In the supplementary material we show that these correlations give rise to a center-of-mass dependent phonon dephasing, which in the Born-Markov approximation reads 
$\Gamma^\mu_\mathbf{Q}=\pi \sum_{\nu \mathbf{q}, \pm} |D^{\mu\nu}_\mathbf{q}|^2 \eta^{\pm}_\mathbf{q} \delta(E^\mu_\mathbf{Q}\pm \Omega_\mathbf{q}-E^\nu_\mathbf{Q+q}).
$
Moreover, additional correlations  $(\cdot)\rvert^\text{rad}_\text{deph.}$ induced by the exciton-photon interaction yield a radiative dephasing for exciton states within the light cone
$\gamma^\mu_\mathbf{\vecPar{k}}=\pi \sum_{\mathbf{k}'} |M^{\mu}_{\mathbf{k}'}|^2 \delta_{\vecPar{k'}\vecPar{k}} \delta(E^\mu_\vecPar{k}-\omega_{\mathbf{k}'}).
$

After the initial thermalization of hot excitons, the occupation numbers $\eta$ and $N$ change slowly compared to the dephasing times and the derived set of equations can be solved in the adiabatic limit (cf. supplementary material).
 It is important to note that a straightforward numerical evaluation of the obtained equations can yield unphysical negative PL signals, when dephasing and density scattering contributions are treated in different approximations.  It is crucial to prevent double counting of the appearing processes as dephasing and as phonon-assisted decay. A detailed discussion is provided in the supplementary material. 
For a thermalized exciton distribution, we find the following analytic expression for the $\sigma$-polarized photon flux emitted in perpendicular direction with respect to the monolayer:
\begin{widetext}
\begin{eqnarray} \label{eq:PL} 
 I_\sigma(\omega)=\dfrac{2}{\hbar}\sum_\mu \dfrac{|M^\mu_{\sigma}|^2}{(E^\mu_\mathbf{0}-\omega)^2+(\gamma^\mu_{\sigma\mathbf{0}}+\Gamma^\mu_\mathbf{0})^2} \bigg{(}  \gamma^\mu_{\sigma\mathbf{0}} N^\mu_\mathbf{0} + \sum_{\nu\mathbf{q},\alpha\pm} |D^{\mu\nu}_{\alpha\mathbf{q}}|^2 N^\nu_\mathbf{q} \eta^{\pm}_{\alpha\mathbf{q}} \dfrac{\Gamma^\nu_\mathbf{q}}{(E^\nu_\mathbf{q}\pm\Omega^\alpha_\mathbf{q} -\omega)^2 +(\Gamma^\nu_\mathbf{q})^2} \bigg{)}.
\end{eqnarray}
\end{widetext}
The first term is analogue to the markovian result for the direct recombination, with the small difference, that the phonon dephasing here only appears in the denominator. However, for energies $\omega\approx E_0$ the second term containing the scattering contributions, can be rewritten as the phonon-guided in-scattering to the bright state, which in thermal equilibrium is equal to the out-scattering $2 \Gamma^\mu_\mathbf{0}N^\mu_\mathbf{0}$. Therefore, in the resonant case, Eq. \ref{eq:PL} can be well approximated with the Elliot formula \cite{koch2006semiconductor}
\begin{eqnarray} \label{eq:Markov}
I_\sigma(\omega)\arrowvert_{\omega\approx E_0}=\dfrac{2}{\hbar}\sum_\mu \dfrac{|M^\mu_{\sigma}|^2(\gamma^\mu_{\sigma\mathbf{0}}+\Gamma^\mu_\mathbf{0})}{(E^\mu_\mathbf{0}-\omega)^2+(\gamma^\mu_{\sigma\mathbf{0}}+\Gamma^\mu_\mathbf{0})^2} N^\mu_\mathbf{0}.
\end{eqnarray}

The second part in Eq. \ref{eq:PL} containing phonon-assisted decay channels agrees with luminescence formulas derived in previous studies \cite{feldtmann2009phonon, feldtmann2010theoretical} for optical phonon replicas in the polaron picture. However, it is here generalized to arbitrary phonon modes, intervalley scattering and a consistent description of the phonon-induced dephasing of direct and indirect transitions.  Note, that the phonon-assisted recombination corresponds to additional luminescence decay channels involving exciton densities of momentum dark states. In contrast, in absorption experiments, phonon-sidebands arise from a frequency dependent dephasing and energy renormalization, which asymmetrically shape the optical response function of the bright exciton \cite{christiansen2017phonon}. The above presented procedure can in principle also be applied to obtain transition rates for other crossed interactions, such as defect-scattering assisted tunnelling.\\


\textit{Phonon-assisted photoluminescence in TMDs:} Now, we apply Eq. \ref{eq:PL} to study the luminescence of TMD monolayers encapsulated with hBN. To this end, we use ab initio parameters from literature for the electronic bandstructure, phonon dispersion, dielectric constants and electron-phonon coupling \cite{kormanyos2015k, jin2014intrinsic, laturia2018dielectric}. The excitonic properties of the system are derived in effective mass approximation by solving the Wannier equation \cite{koch2006semiconductor}. Details about the used parameters, the applied screening model and the form of the excitonic matrix elements are given in the supplementary material.

\begin{figure}[b!]
\includegraphics[width=80mm]{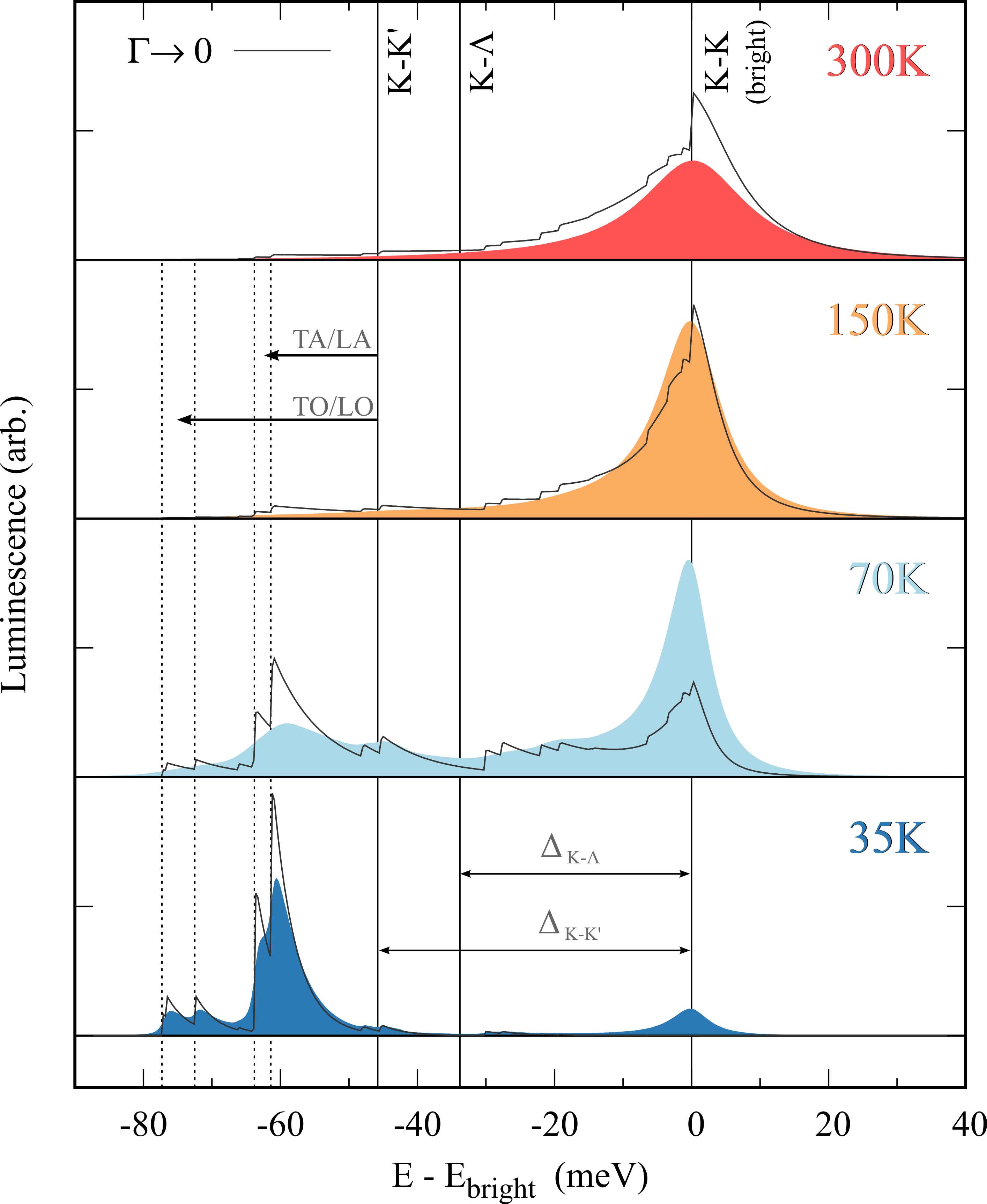}
\caption{PL spectra of hBN-encapsulated WSe$_2$ monolayers at four different temperatures. Together with the full simulation (colored curves) we also show the corresponding Fermi's golden rule solution (thin solid lines). At higher temperatures, the lower lying momentum-dark states K-K' and K-$\Lambda$ give rise to a strong asymmetric broadening of the bright exciton resonance towards lower energies. For low temperatures, the PL signal exhibits indirect peaks stemming from the phonon-assisted recombination of dark excitons.}
\label{fig:Tstudy} 
\end{figure}

Figure \ref{fig:Tstudy} shows calculated PL spectra for hBN-encapsulated monolayer WSe$_2$ at four different temperatures.
In addition to the full spectra calculated from Eq. \ref{eq:PL} (color shaded), we also present the spectra in the case of $\Gamma^\nu_\mathbf{q}\longrightarrow 0$, corresponding to the Fermi's golden rule (solid line). The sharp steps in the solid line resulting from the strict energy conservation in the Fermi's golden rule conveniently illustrate the multitude of exciton valleys and phonon modes contributing to the overall PL signal. Here, each step in the spectrum corresponds to a transition from the energetic minimum of a certain valley while absorbing/emitting a phonon.

At 300K and 150K the PL spectrum is dominated by the bright exciton resonance, but already here phonon-assisted indirect recombinations have a significant impact resulting in an asymmetric resonance. At energies larger than $E_0$ the signal is mainly shaped by recombinations of small momentum excitons in the K-K valley, which scatter to virtual states in the light cone via low energy acoustic phonons. For energies smaller then $E_0$, much larger broadening is observed. This can be ascribed to K-K excitons emitting optical phonons and - more importantly - to intervalley transitions from the K-$\Lambda$ and the K-K' state (hole located at the K and  electron at the $\Lambda$ and the K' valley, respectively). For the considered dielectric environment our Wannier model predicts that the latter are located about $\unit[34]{meV}$ and $\unit[46]{meV}$ below the bright exciton, respectively. 

 At low temperatures, the optical response shifts towards multiple indirect PL peaks below the bright exciton stemming from the phonon-assisted recombination of the energetically lowest K-K' exciton. Although scattering to virtual states becomes more improbable with increasing detuning from the bright state, the large population ratio between dark and bright states gives rise to a strong indirect signal at low temperatures  (Fig. \ref{fig:scheme}). The predicted low temperature PL signals at about $-\unit[60]{meV}$ (K-K' - acoustic phonon assisted) and $-\unit[75]{meV}$ (K-K' - optical phonon assisted) correspond well to experimentally observed PL peaks below the trion resonance. In agreement with our theory, these peaks are not visible in reflection/absorption spectra \cite{christiansen2017phonon}. Although clearly visible in PL spectra, so far these peaks were to a large extent ignored in literature due to their unclear origin. Our work reveals that these peaks stem from indirect phonon-assisted transitions from lower lying momentum-dark excitons in tungsten-based TMDs. We have also performed calculations for WS$_2$ (not shown) obtaining similar low-temperature PL features.
 
 \begin{figure}[t!]
\includegraphics[width=80mm]{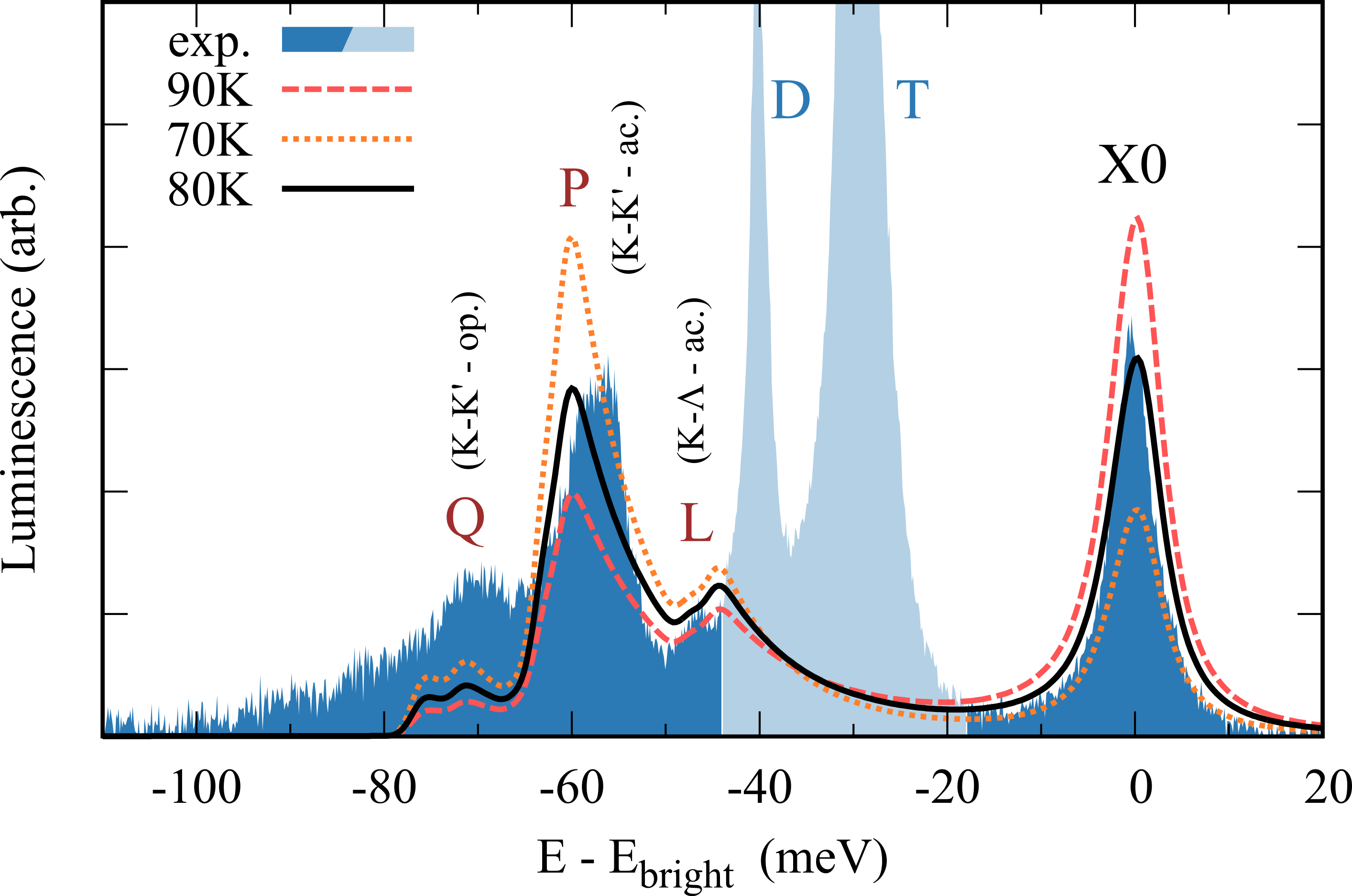}
\caption{Direct comparison between experiment and theory. The blue shaded curve shows the PL spectrum measured on hBN-encapsulated WSe$_2$ at charge neutrality at T=15~K. Lines show simulated spectra for 15K lattice temperature and three different exciton temperatures.}
\label{fig:exp} 
\end{figure}

In Fig. \ref{fig:exp} we directly compare an experimentally measured PL spectrum with our simulation. The blue shaded curve shows a spectrum measured at T=15~K for hBN-encapsulated WSe$_2$ at charge neutrality. Details about the experiment can be found in Ref. \cite{courtade2017charged}. The solid and dashed lines show calculated spectra for 15K phonon temperature and three different effective exciton temperatures. For a better comparisson the calculated spectra where convoluted with a 1 meV broad Gaussian to mimic weak disorder. Note that the calculated spectra in Fig. \ref{fig:Tstudy} correspond to stationary PL signals of exciton distributions which are in thermal equilibrium with the lattice. In Fig. \ref{fig:exp} we introduce different phonon and exciton temperatures to take into account that in experiments performed at low temperatures the optically injected hot excitons decay before they have reached thermal equilibrium with the lattice.  
Figure \ref{fig:exp} illustrates that the theoretically predicted luminescence features from dark excitons agree well with the experimentally observed resonances between 45 and 80meV below the bright exciton denoted with L, P and Q. The two pronounced peaks around 40 and 30 meV below the bright state (denoted with D and T) potentially stem from spin-forbidden states and residual trions \cite{ye2018efficient}, which are not included in our model. We find the best agreement between experiment and theory for an effective exciton temperature of about 80K, indicating that the luminescence stems from an exciton distribution that is much hoter than the lattice. In the supplementary material we have summarized several independent PL measurements on hBN-encapsulated WSe$_2$ \cite{courtade2017charged, ye2018efficient,barbone2018charge} in which the above discussed phonon-assisted peaks have been observed. It is important to note, that the peak denoted as P in Fig. \ref{fig:exp}, which we attribute to the acoustic phonon assisted decay of K-K' excitons, in experiments performed at 4~K shows a double peak structure. In the supplemetary, we have included a discussion about this splitting.

\begin{figure}[t!]
\includegraphics[width=85mm]{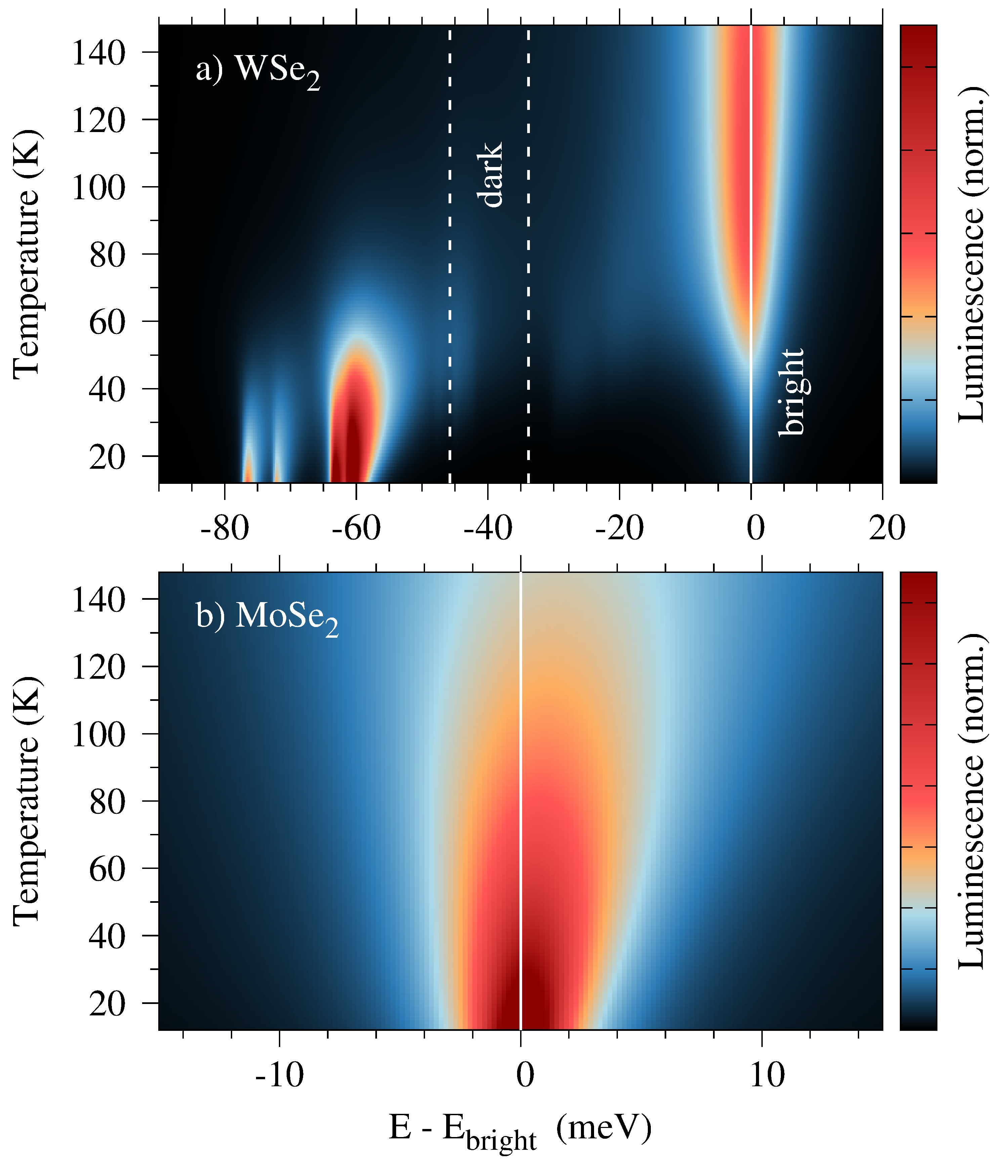}
\caption{Temperature-dependent PL spectra calculated for (a) WSe$_2$ and (b) MoSe$_2$. For each temperature the spectra have been normalized to the integrated PL. While WSe$_2$ shows a clearly asymmetric broadening towards lower energies and distinct indirect peaks at low temperatures,  MoSe$_2$ has no additional indirect peaks and is stronger broadened towards higher energies.}
\label{fig:Tscan} 
\end{figure}

Finally, we compare our results obtained for WSe$_2$ (representing tungsten-based 2D materials) with molybdenum-based TMDs, cf. Fig. \ref{fig:Tscan}.  
Here, we show a continuous temperature study of the PL spectrum for (a) WSe$_2$ and (b) MoSe$_2$. In tungsten-based TMDs the lower lying K-$\Lambda$ and K-K' valley provide a continuous density of (momentum-dark) states below the bright exciton, yielding an asymmetric broadening of the bright exciton peak towards lower energies. In contrast, for MoSe$_2$ the bright state constitutes the global minimum of the exciton dispersion calculated in our Wannier model. Therefore there are no phonon-assisted PL peaks below the bright state, and the main peak is asymmetrically broadened towards higher energies, cf. Fig. \ref{fig:Tscan}(b). Here the broadening of the low energy side of the main line is predominantly given by the energy uncertainty of the bright state giving rise to a Lorentzian shape. In contrast, the high energy side is additionally shaped by the decay of small momentum excitons in the K-K valley assisted by long range acoustic phonons, which agrees with the findings in Ref. \onlinecite{shree2018observation}.

In summary, we have presented an analytical expression for the phonon-assisted exciton photoluminescence from momentum-dark excitons which allows to model the emission spectrum of an arbitrary semiconducting material. When applying our model to the luminescence of WSe$_2$, we find in good agreement with experiment clear PL signals stemming from dark intervalley excitons. This explains the origin of so far observed but unidentified PL peaks. In contrast, MoSe$_2$ does not exhibit indirect PL peaks and shows an opposite asymmetric broadening which is consistent with the absence of lower lying dark exciton states. Our work will trigger future experimental studies on dark exciton PL and in particular time-resolved PL measurements addressing exciton thermalization.

\begin{acknowledgments}
The Chalmers group acknowledges financial support from the European Unions Horizon 2020 research and innovation program under grant agreement No 785219 (Graphene Flagship) as well as from the Swedish Research Council (VR). The TUB group was funded by the Deutsche Forschungsgemeinschaft (DFG) vi project number 182087777 in SFB 951 (project B12, D.C., M.S and A.K.) and by the European Unions Horizon 2020 research and innovation program under Grant Agreements No. 734690 (SONAR, A.K.). F.K. and D.C. thank the School of Nanophotonics (SFB 787) for financial support. C.R. and B.U. acknowledge funding from ANR 2D-vdW-Spin, ANR VallEx, Labex NEXT projects VWspin and MILO, ITN Spin-NANO Marie Sklodowska-Curie grant agreement No 676108 and ITN 4PHOTON Nr. 721394.
\end{acknowledgments}

\onecolumngrid
\appendix
\newpage
\noindent \begin{center}
\textbf{\large Phonon-assisted Photoluminescence from Dark Excitons in Monolayers of Transition Metal Dichalcogenides   \\ 
\vspace*{5mm}
--SUPPLEMENTARY MATERIAL--}
\par\end{center}

\begin{center}
Samuel Brem$^{1*}$, August Ekman$^{1}$, Dominik Christiansen$^2$, Florian Katsch$^2$, Malte Selig$^2$, Cedric Robert$^3$, Xavier Marie$^3$, Bernhard Urbaszek$^3$, Andreas Knorr$^2$, Ermin Malic$^{2}${\small }\\
{\small $^{1}$ }\textit{\small Chalmers University of Technology, Department of Physics, 41296 Gothenburg, Sweden}\\
{\small $^{2}$ }\textit{\small Technical University Berlin, Institute of Theoretical Physics, 10623 Berlin, Germany}\\
{\small $^{3}$ }\textit{\small Universit\'e de Toulouse, INSA-CNRS-UPS, LPCNO, 31077 Toulouse, France}
\par\end{center}{\small \par}

\section{Equations of Motion}

 In the low excitation regime the fermionic substructure of excitons as well as the exciton-exciton interaction can be neglected, so that the properties of the TMD monolayer are described by the following excitonic hamiltonian in rotating frame:

\begin{eqnarray} \label{eq:Hamilton}
H&=&\sum_{\mu\vecPar{Q}}  E^{\mu}_{\vecPar{Q}} a^{\dagger}_{\mu \vecPar{Q}}a^{\phantom\dagger}_{\mu \vecPar{Q}} + \sum_{\alpha\vecPar{q}}\Omega^{\alpha}_{\vecPar{q}} b^{\dagger}_{\alpha \vecPar{q}}b^{\phantom\dagger}_{\alpha \vecPar{q}} + \sum_{\sigma\mathbf{k}}\omega^{\sigma}_{\mathbf{k}} c^{\dagger}_{\sigma \mathbf{k}}c^{\phantom\dagger}_{\sigma \mathbf{k}} \nonumber\\ 
  &+& \sum_{\mu\sigma \mathbf{k}} (M^{\mu}_{\sigma \mathbf{k}} c^{\dagger}_{\sigma \mathbf{k}}a^{\phantom\dagger}_{\mu \vecPar{k}}+M^{\mu\ast}_{\sigma \mathbf{k}} a^{\dagger}_{\mu \vecPar{k}} c^{\phantom\dagger}_{\sigma \mathbf{k}}) \nonumber \\
  &+& \sum_{\nu\mu  \vecPar{Q} \alpha \vecPar{q}}D^{\nu\mu}_{\alpha \vecPar{q}} a^{\dagger}_{\nu \vecPar{Q}+\vecPar{q}}a^{\phantom\dagger}_{\mu \vecPar{Q}}(b^{\phantom\dagger}_{\alpha \vecPar{q}}+b^{\dagger}_{\alpha, -\vecPar{q}}).
\end{eqnarray}
For definitions see main text. For convinience we will in the following supress the photon, phonon and exciton mode index. The derivation shown below can be performed analogously for multiple exciton states and photon/phonon modes, yielding and additive solution. To further reduce the amount of indices in the equations we consider the case of perpendicular emission to the monolayer plane, removing the in-plane momentum transfer due to interaction with photons, whose consideration is however straight forward.
Applying the Heisenberg equation of motion $i\hbar\partial_t\langle\mathcal{\cdot}\rangle= \langle [\cdot,H]\rangle$ to calculate the evolution of the photon density $n_{\mathbf{k}}=\langle  c^{\dagger}_{\mathbf{k}}c^{\phantom\dagger}_{\mathbf{k}}\rangle$ we find a coupled system of equations involving the polarization $\mathcal{S}_\mathbf{k}=\langle c^{\dagger}_{\mathbf{k}}a^{\phantom\dagger}_{0}\rangle$, the phonon-assisted polarization $\mathcal{U}^{\pm}_{\mathbf{kq}}=\langle c^{\dagger}_{\mathbf{k}}b^{(\dagger)}_{\mp\mathbf{q}}a^{\phantom\dagger}_{-\mathbf{q}}\rangle$ (with phonon creation for the $+$ case), the exciton-phonon correlation $\mathcal{C}^{\pm}_{\mathbf{Qq}}=\langle a^{\dagger}_{\mathbf{Q}} a^{\phantom\dagger}_{\mathbf{Q}-\mathbf{q}}b^{(\dagger)}_{\mp\mathbf{q}}\rangle$  and the exciton density $N_{\mathbf{Q}}=\langle a^{\dagger}_{ \mathbf{Q}}a^{\phantom\dagger}_{\mathbf{Q}}\rangle$:
\begin{eqnarray} 
\dfrac{d}{dt} n_{\mathbf{k}} &=& \dfrac{2}{\hbar}\Im\text{m}\{M_{\mathbf{k}}\mathcal{S}_\mathbf{k}\}, \label{eq:EoM_n} 
\\
i \hbar \dfrac{d}{dt}\mathcal{S}_\mathbf{k} &=& (E_{0}-\omega_{\mathbf{k}})\mathcal{S}_\mathbf{k} - M^{\ast}_{\mathbf{k}}N_{0} +\sum_{\mathbf{q},\zeta=\pm} D_{\mathbf{q} }\mathcal{U}^{\zeta}_{\mathbf{kq}} +\sum_{\mathbf{k'}} M^{\ast}_{\mathbf{k'}} n_{\mathbf{kk'}}\\
i \hbar \dfrac{d}{dt}  \mathcal{U}^{\zeta}_{\mathbf{kq}} &=& (E_{\mathbf{-q}}-\zeta\Omega_\mathbf{q} -\omega_\mathbf{k}) \mathcal{U}^{\zeta}_{\mathbf{kq}} - M^{\ast}_{\mathbf{k}} \mathcal{C}^{\zeta}_{\mathbf{0q}} + D^{\ast}_{\mathbf{q}} \eta^{\zeta}_\mathbf{q} \mathcal{S}_\mathbf{k} +\sum_{\mathbf{q'},\zeta'=\pm}  D_{\mathbf{q'}}X^{\zeta\zeta'}_{\mathbf{kqq'}}, \label{eq:Udot}
\\
i \hbar \dfrac{d}{dt} \mathcal{C}^{\zeta}_{\mathbf{Qq}} &=& (E_{\mathbf{Q}-\mathbf{q}}-\zeta\Omega_\mathbf{q}-E_{\mathbf{Q}}) \mathcal{C}^{\zeta}_{\mathbf{Q}\mathbf{q}}- D^{\ast}_{\mathbf{q}}Q^{\zeta}_{\mathbf{Q}\mathbf{q}} \nonumber\\
&-&\sum_{\mathbf{k}} (M_{\mathbf{k}}\mathcal{U}^{\zeta}_{\mathbf{kq}}\delta_{\mathbf{Q},0}-M_{\mathbf{k}}^\ast\mathcal{U}^{-\zeta}_{\mathbf{k,-q}}\delta_{\mathbf{Q},-\mathbf{q}}) +\sum_{\mathbf{q'},\zeta'=\pm} (D_{\mathbf{q'}}Y^{\zeta\zeta'}_{\mathbf{Qqq'}}- D^\ast_{\mathbf{q'}}Y^{\zeta\zeta'}_{\mathbf{Q+q',q,-q'}}), \label{eq:Cdot} \\
\dfrac{d}{dt} N_{\mathbf{Q}} &=& -\dfrac{2}{\hbar}\sum_\mathbf{k}\Im\text{m}\{M_{\mathbf{k}}\mathcal{S}_\mathbf{k}\}\delta_{\mathbf{Q,0}}-\dfrac{i}{\hbar}\sum_{\mathbf{q},\zeta=\pm}D_\mathbf{q}(\mathcal{C}^{\zeta}_{\mathbf{Qq}}-\mathcal{C}^{\zeta}_{\mathbf{Q+q,-q}})
\end{eqnarray} 
Here $\eta^{\pm}_\mathbf{q}=1/2\mp 1/2 +\langle b^{\dagger}_{\mathbf{q}} b^{\phantom\dagger}_{\mathbf{q}}\rangle$ denotes the relevant phonon occupation factor for absorbtion/emission, which are assumed to be time independent (bath approximation). Moreover, $Q^{\pm}_{\mathbf{Q}\mathbf{q}}=\eta^{\mp}_\mathbf{q} N_{\mathbf{Q}-\mathbf{q}}-\eta^{\pm}_\mathbf{q} N_{\mathbf{Q}}$ is the source of the corresponding exciton-phonon correlation and $n_{\mathbf{kk'}}=\langle c^{\dagger}_{\mathbf{k}}c^{\phantom\dagger}_{\mathbf{k'}}\rangle$ stands for the photonic density matrix. 

To obtain the above set of equations we have factorized appearing many-particle expectation values acording to the cluster expansion scheme [cite], viz. for example  
\begin{eqnarray} 
\langle c^{\dagger}_{\mathbf{k}}b^{\dagger}_{\mathbf{q}}b_{\mathbf{q'}}a_{\mathbf{q}-\mathbf{q'}}\rangle\approx\langle c^{\dagger}_{\mathbf{k}}a_{\mathbf{q}-\mathbf{q'}}\rangle\langle b^{\dagger}_{\mathbf{q}}b_{\mathbf{q'}}\rangle+\langle c^{\dagger}_{\mathbf{k}}b^{\dagger}_{\mathbf{q}}b_{\mathbf{q'}}a_{\mathbf{q}-\mathbf{q'}}\rangle^\text{corr.}
\end{eqnarray} 
Which means that we neglect coherent quantities, e.g. $\langle a\rangle$, assuming a fast decoherence after the initial laser excitation, and discard pure photon-phonon correlations. Moreover, we have introduced the notation $\langle\cdot\rangle^\text{corr.}$, which in the upper case quantifies the correlation between phonon density and polarization. These higher order correlations are abbreviated in Eq. \ref{eq:Udot} and \ref{eq:Cdot} via $X^{\zeta\zeta'}_{\mathbf{kqq'}}=\langle c^{\dagger}_{\mathbf{k}}b^{(\dagger)}_{-\zeta\mathbf{q}}b^{(\dagger)}_{-\zeta'\mathbf{q}'}a^{\phantom\dagger}_{-\mathbf{q}-\mathbf{q'}}\rangle^\text{corr.}$ and  $Y^{\zeta\zeta'}_{\mathbf{Qqq'}}=\langle b^{(\dagger)}_{-\zeta\mathbf{q}}b^{(\dagger)}_{-\zeta'\mathbf{q}'}a^{\dagger}_{\mathbf{Q}}a^{\phantom\dagger}_{\mathbf{Q}-\mathbf{q}-\mathbf{q'}}\rangle^\text{corr.}$, while the case $\zeta=+1$ again corresponds to the phonon creation. Similar as the coupling of $\mathcal{S}$ to $\mathcal{U}$ gives rise to a phonon induced energy renormalization as well as additional phonon assisted transitions, the coupling of $\mathcal{U}$ to $X$ introduces two-phonon-assisted transitions as well as an energy renormalization for $\mathcal{U}$. The microscopic treatment of higher order phonon-emission sidebands goes beyond the scope of this work, however, we will include the energy renormalizations induced by the 2-phonon-correlations.

\section{Many-Particle Dephasing}

To obtain the many-particle dephasing induced by electron-phonon interaction, we additionally consider the equations of motion of $X$ and $Y$.  In the following we only show the derivation for $X$, since  $Y$ can be treated in complete analogy. Applying the same approximations as above we find:

\begin{eqnarray} 
i \hbar \dfrac{d}{dt} X^{\zeta\zeta'}_{\mathbf{kqq'}} =(E_{\mathbf{q+q'}}-\zeta\Omega_\mathbf{q} -\zeta'\Omega_\mathbf{q'} -\omega_\mathbf{k})X^{\zeta\zeta'}_{\mathbf{kqq'}} +D^{\ast}_{\mathbf{q}'} \eta^{\zeta'}_\mathbf{q'} \mathcal{U}^{\zeta}_{\mathbf{kq}} + \dot{X}\rvert_\text{source} +\dot{X}\rvert_\text{corr.}.
\end{eqnarray} 

Additional source terms $\dot{X}\rvert_\text{source}$ following from the exciton-photon interaction give rise to two-phonon-assisted photoemission processes, which will not be considered here. Moreover,  higher order exciton-phonon correlations $\dot{X}\rvert_\text{corr.}$ give rise to equivalent energy renormalizations as $X$ induces for $\mathcal{U}$, denoted with $\delta^{+}$ which will be taken into accoutn below via a self-consistent renormalization. In the static limit we obtain,
\begin{eqnarray} \label{eq:MarkovPi}
 X^{\zeta\zeta'}_{\mathbf{kqq'}} \approx  -\dfrac{D^{\ast}_{\mathbf{q}'} \eta^{\zeta'}_\mathbf{q'} \mathcal{U}^{\zeta}_{\mathbf{kq}}}{E_{\mathbf{q+q'}}-\zeta\Omega_\mathbf{q} -\zeta'\Omega_\mathbf{q'} -\omega_\mathbf{k} -i\delta^{+} } \approx  -\dfrac{D^{\ast}_{\mathbf{q}'} \eta^{\zeta'}_\mathbf{q'} \mathcal{U}^{\zeta}_{\mathbf{kq}}}{E_{\mathbf{q+q'}}-E_{\mathbf{q}} -\zeta'\Omega_\mathbf{q'} -i\delta^{+}} ,
\end{eqnarray} 
where in the second step we assume that $\mathcal{U}^{\zeta}_{\mathbf{kq}}$ only gives a non-zero response for photon energies close to its resonance, viz. $\omega_\mathbf{k}\approx E_{\mathbf{q}}-\zeta\Omega_\mathbf{q}$ which is analogous to the treatment of scattering contributions in second order Born-Markov-approximation [cite]. Plugging Eq. \ref{eq:MarkovPi} into \ref{eq:Udot} , we find that the higher order exciton-phonon correlations give rise to the polaronic energy renormalization
\begin{eqnarray} \label{eq:Renorm}
 E_\mathbf{Q} \longrightarrow\tilde{E}_\mathbf{Q}=E_\mathbf{Q} + \Sigma^\text{ph}_\mathbf{Q}; \;\;\;\;\;\;\;  \Sigma^\text{ph}_\mathbf{Q}=\sum_{\zeta\mathbf{q}}\dfrac{\lvert D_\mathbf{q}\rvert^2 \eta^{\zeta}_\mathbf{q}}{\tilde{E}_{\mathbf{Q}}-\tilde{E}_{\mathbf{Q+q}} +\zeta\Omega_\mathbf{q}},
\end{eqnarray} 
Where $\delta^{+}$ is assumed to yield the same energy renormalization for $X$ as for all other orders, giving rise to the self-consistent polaron self-energy $\Sigma^\text{ph}$. The same treatment of $Y$ yields an analogous result in Eq. \ref{eq:Cdot}. Moreover, the coupling of $\mathcal{S}$ to $n_{\mathbf{kk'}}$ and the backcoupling of $\mathcal{C}$ to $\mathcal{U}$ can be similarly shown to yield a radiative dephasing for exciton states in the lightcone,
\begin{eqnarray}
 E_\mathbf{0} \longrightarrow E_\mathbf{0} - i\gamma_0; \;\;\;\;\;\;\;  \gamma_0=\dfrac{1}{i}\sum_{\mathbf{k}}\dfrac{\lvert M_\mathbf{k}\rvert^2}{E_{0}-\omega_{\mathbf{k}}-i\kappa }\approx \pi\sum_{\mathbf{k}}\lvert M_\mathbf{k}\rvert^2 \delta(E_{0}-\omega_{\mathbf{k}}),
\end{eqnarray} 
where we assume short photon lifetimes $1/\kappa$ and a weakly varying coupling element $M_\mathbf{k}$ in vicinity of $E_0$.

\section{Photoluminescence in Static Limit}
Using the derived renormalizations, $E_\mathbf{Q} \longrightarrow\tilde{E}_\mathbf{Q}=E_\mathbf{Q} + \Sigma^\text{ph}_\mathbf{Q} -i\gamma_\mathbf{0}\delta_{\mathbf{Q,0}}$, we now solve Eq. \ref{eq:EoM_n} - \ref{eq:Cdot} by assuming slowly varying densities allowing to find adiabatic solutions from the static limit of our equations of motion. Introducing the Greensfunctions 
\begin{eqnarray} \label{eq:Green}
G_\mathbf{Q}(\omega)=(E_\mathbf{Q}-\omega)^{-1}; \;\;\;\;\;\;\;
\mathcal{G}_\mathbf{Q}(\omega)=(\tilde{E}_{\mathbf{Q}}-\omega)^{-1},
\end{eqnarray}
we find:
\begin{eqnarray} \label{eq:step}
\mathcal{S}_\mathbf{k}&=&G_0(\omega_{\mathbf{k}}+i\gamma_\mathbf{0})\bigg( M^\ast_\mathbf{k} N_0+\sum_{\zeta\mathbf{q}}\mathcal{G}_\mathbf{q}(\omega_\mathbf{k}+\zeta\Omega_\mathbf{q})\lvert D_\mathbf{q}\rvert^2 \eta^{\zeta}_\mathbf{q} \mathcal{S}_\mathbf{k}\bigg) \nonumber\\
&-& G_0(\omega_{\mathbf{k}}+i\gamma_\mathbf{0}) M^\ast_\mathbf{k}\sum_{\zeta\mathbf{q}}\mathcal{G}_\mathbf{q}(\omega_\mathbf{k}+\zeta\Omega_\mathbf{q}) \mathcal{G}_\mathbf{q}(\tilde{E}^\ast_0+\zeta\Omega_\mathbf{q}) \lvert D_\mathbf{q}\rvert^2 Q^{\zeta}_{\mathbf{0}\mathbf{q}}
\end{eqnarray}
Now we can identify,
\begin{eqnarray} \label{eq:Trick1}
\Sigma^\text{ph}(\omega)=\sum_{\zeta\mathbf{q}}\mathcal{G}_\mathbf{q}(\omega+\zeta\Omega_\mathbf{q})\lvert D_\mathbf{q}\rvert^2 \eta^{\zeta}_\mathbf{q}  \;\;\;\;\;\;\; \text{and} \;\;\;\;\;\;\; \dfrac{G_0(\omega+i\gamma_\mathbf{0})}{1+\Sigma^\text{ph}(\omega)G_0(\omega+i\gamma_\mathbf{0})} \approx  \mathcal{G}_\mathbf{0}(\omega),
\end{eqnarray}
where we consistent with Eq. \ref{eq:MarkovPi} evaluate the selfenergy at resonance. Next we recast
\begin{eqnarray} \label{eq:Trick2}
\mathcal{G}_\mathbf{q}(\omega+\zeta\Omega_\mathbf{q}) \mathcal{G}_\mathbf{q}(\tilde{E}^\ast_0+\zeta\Omega_\mathbf{q})=-\mathcal{G}^\ast_\mathbf{0}(\omega)\bigg(\mathcal{G}_\mathbf{q}(\omega+\zeta\Omega_\mathbf{q})- \mathcal{G}_\mathbf{q}(\tilde{E}^\ast_0+\zeta\Omega_\mathbf{q})\bigg)
\end{eqnarray}

Hence, pluging Eq.\ref{eq:step} in \ref{eq:EoM_n} yields,
\begin{eqnarray} \label{eq:prePL}
\dfrac{d}{dt} n_\mathbf{k} = \dfrac{2}{\hbar} |M_\mathbf{k}|^2 &\bigg(&N_0 \Im\text{m}\{\mathcal{G}_0(\omega_\mathbf{k})\} \nonumber\\
 &+& \sum_{\mathbf{q},\zeta} |D_{\mathbf{q}}\mathcal{G}_0(\omega_\mathbf{k})|^2 (\eta^{-\zeta}_\mathbf{q} N_{-\mathbf{q}}-\eta^{\zeta}_\mathbf{q} N_{0}) \Im\text{m}\{\mathcal{G}_\mathbf{q}(\omega_\mathbf{k}+\zeta\Omega_\mathbf{q})-\mathcal{G}_\mathbf{q}(\tilde{E^\ast_0}+ \zeta\Omega_\mathbf{q})\}\bigg), 
\end{eqnarray}

When considering the equation of motion for the exciton density in the lightcone, the phonon-scattering contribution in the static limit can be writen as
\begin{eqnarray} \label{eq:Trick3}
\dfrac{d}{dt} N_{\mathbf{0}} \arrowvert_\text{phon.}&=& -\dfrac{i}{\hbar}\sum_{\mathbf{q},\zeta=\pm}D_\mathbf{q}(\mathcal{C}^{\zeta}_{\mathbf{0q}}-\mathcal{C}^{\zeta}_{\mathbf{q,-q}}) \nonumber\\
&=&\dfrac{2}{\hbar}\sum_{\mathbf{q},\zeta} |D_{\mathbf{q}}|^2 (\eta^{-\zeta}_\mathbf{q} N_{-\mathbf{q}}-\eta^{\zeta}_\mathbf{q} N_{0}) \Im\text{m}\{\mathcal{G}_\mathbf{q}(\tilde{E^\ast_0}+ \zeta\Omega_\mathbf{q})\}\approx 0
\end{eqnarray}
resembling the semi-classical Boltzman scattering-equations. This temporal change of the exciton occupation exactly corresponds to the second contribution in the second line of Eq. \ref{eq:prePL}. Since we have assumed slowly varying occupations we have to neglect this contribution to stay consisent with our adiabatic solution.

Finally, the contribution proportional to $N_0$ of the remaining term in the second line of Eq. \ref{eq:prePL}  can again acording to Eq. \ref{eq:Trick1} be rewritten as $-|\mathcal{G}_0(\omega_\mathbf{k})|^2 \Im\text{m}\{\Sigma^\text{ph}(\omega_\mathbf{k})\}N_0$, which cancels the imaginary part of the polaron self-energy in the numerator of the first line in Eq. \ref{eq:prePL}. Throughout this work we neglect polaron shifts and only consider the imaginary part of the self-energy (dephasing),
\begin{eqnarray} 
\Sigma^\text{ph}_\mathbf{Q}\approx -i\Gamma_\mathbf{Q}=-i\pi \sum_{ \mathbf{q}, \zeta} |D_\mathbf{q}|^2 \eta^{\zeta}_\mathbf{q} \delta(E_\mathbf{Q}-E_\mathbf{Q+q}+\zeta\Omega_\mathbf{q}),
\end{eqnarray} 
Hence, we obtain
\begin{eqnarray} \label{eq:PL} 
\dfrac{d}{dt} n_\mathbf{k} =\dfrac{2}{\hbar} \dfrac{|M_\mathbf{k}|^2}{(E_\mathbf{0}-\omega_\mathbf{k})^2+(\gamma_{\mathbf{0}}+\Gamma_\mathbf{0})^2} \bigg{(}  \gamma_{\mathbf{0}} N_\mathbf{0} + \sum_{\mathbf{q}\zeta} |D_{\mathbf{q}}|^2 N_\mathbf{q} \eta^{\zeta}_{\mathbf{q}} \dfrac{\Gamma_\mathbf{q}}{(E_\mathbf{q}+\zeta\Omega_\mathbf{q} -\omega_\mathbf{k})^2 +(\Gamma_\mathbf{q})^2} \bigg{)}         
\end{eqnarray}
The generalization to several exciton,phonon and photon modes is straight forward and can be found in the main text.

\section{Excitonic Matrix Elements and Wavefunctions}

To evaluate the above derived PL formula we make use of the Mott-Wannier model of excitons. In this framework the exciton-phonon matrix element is determined by the electron-phonon couplings for electrons and holes $g^{e/h}$ and the exciton wavefunction in momentum space $\Phi(\mathbf{k})$ via \cite{selig2018dark,brem2018exciton}
\begin{eqnarray} \label{eq:ex-ph}
D^{\nu \mu}_{\lambda\mathbf{q}}=\sum_{\mathbf{k},\alpha=e,h} g^{\alpha\mathbf{k}}_{\lambda \mathbf{q}}\,\Phi_{\nu}^{\ast}(\mathbf{k}) \Phi_\mu(\mathbf{k+q_\alpha}).
\end{eqnarray} 
Here the momentum transferred to the electron (hole) is denoted by $q_{e(h)}$, when the exciton gains a center-of-mass momentum $q=q_e+q_h$. The exciton index $\nu$ here acts as a compound index containing principal quantum number, angular momentum, electron/hole valley and spin configuration. For the carrier-phonon coupling $g$ we use the deformation potential approximations for acoustic and optical phonons deduced from density functional perturbation theory (DFPT) in Ref. \cite{jin2014intrinsic}. Furthermore, the radiative dephasing is determined via 
\begin{eqnarray} 
\gamma^\nu_{\sigma 0}= \pi\sum_{\mathbf{k}}\lvert M^\nu_{\sigma\mathbf{k}}\rvert^2 \delta(E^{\nu}_{0}-\omega_{\mathbf{k}})=\dfrac{\hbar e_0^2}{2 m_0^2\epsilon_0 n c_0} |\mathbf{m}^{vc}\cdot \mathbf{e}_\sigma|^2 \dfrac{\lvert\tilde{\Phi}_\nu(r=0)\rvert^2}{E^\nu_0},
\end{eqnarray} 
where the interband momentum matrix element $\mathbf{m}^{vc}$ is derived from a two band $k\cdot p$ Hamiltonian, which in vicinity of the K point yields \cite{xiao2012coupled}
\begin{eqnarray} \label{eq:optical}
|\mathbf{m}^{vc} \cdot \mathbf{e}_\sigma|^2=\dfrac{1}{2}[\dfrac{a_0m_0t}{\hbar}(1+\sigma)]^2.
\end{eqnarray}
The next neighbor hopping integral $t=\hbar/a_0 \sqrt{E_g/(m_e+m_h)}$ is determined by the effective masses $m_{e/h}$ of electrons and holes and the single particle bandgap $E_g$ at the K-point, while $\sigma=\pm 1$ for left-(right-)handed circularly polarized light. To calculate the exciton binding energies in a monolayer we use an approach analog to the Keldish screening for charges in a thin film of thickness d surrounded by a dielectric environment. We explicitly take into account anisotropic dielectric tensors. Solving the Poisson equation for the boundary conditions of an encapsulated monolayer yields $W_q=V_q/\epsilon_{scr}(q)$, with the bare 2D-Fourier transformed Coulomb potental $V_q$ and the non-local screening,
\begin{eqnarray} \label{eq:keldish}
\epsilon_{scr}(q)= \kappa_1 \tanh(\dfrac{1}{2}[\alpha_1dq-\ln(\dfrac{\kappa_1-\kappa_2}{\kappa_1+\kappa_2})]),
\end{eqnarray}
where $\kappa_{i}=\sqrt{ \epsilon^{\parallel}_i \epsilon^{\bot}_i}$ and $\alpha_i=\sqrt{ \epsilon^{\parallel}_i /\epsilon^{\bot}_i}$ account for the parallel and perpendicular component of the dielectric tensor $\epsilon$ of the monolayer ($i=1$) and the environment ($i=2$), which can be found in refs. \cite{laturia2018dielectric, geick1966normal}.

Finally, to calculate the excitonic wavefunctions we numerically solve the the Wannier equation,
\begin{eqnarray} \label{eq:wannier}
 (\varepsilon_{c,\mathbf{k}+\alpha\mathbf{Q}}-\varepsilon_{v,\mathbf{k}-\beta\mathbf{Q}}) \Phi_{\nu\mathbf{Q}}(\mathbf{k}) - \sum_\mathbf{q} W_\mathbf{q} \Phi_{\nu\mathbf{Q}}(\mathbf{k+q}) = E_{\nu\mathbf{Q}} \Phi_{\nu\mathbf{Q}}(\mathbf{k}).
\end{eqnarray}
Here  we use relative (k) and center-of mass coordinates (Q) with $\alpha(\beta)=m_{c(v)}/(m_c+m_v)$ assuming effective electron ($m_c$) and hole ($m_v$) masses. Within the vicinity of minima and maxima of valence and conduction band, we approximate the dispersions quadratically, which allows us to separate relative and center of mass motion. When $\mathbf{K}_c$ denotes the conduction band valley and $\mathbf{K}_v$ the valence band valley, we find $\Phi_{\nu\mathbf{Q}}(\mathbf{k})=\Phi_{\nu}(\mathbf{k})=\Psi_\nu(\mathbf{k}-\alpha \mathbf{K}_v-\beta \mathbf{K}_c)$, with $\Psi$ obeying the effective electron-hole Schroedinger equation,
\begin{eqnarray} \label{eq:wannier2}
\frac{\hbar^2 k^2}{2 m_{\text{r}}} \Psi_\nu(\mathbf{k}) - \sum_\mathbf{q} W_\mathbf{q} \Psi_\nu(\mathbf{k+q}) = E^\text{bind}_\nu\Psi_\nu(\mathbf{k}).
\end{eqnarray}
where $m_\text{r}=(m_c m_v)/(m_c+m_v)$ is the reduced exciton mass for the corresponding valley. Furthermore, the parabolic approximation yields the center-of-mass dispersion $E_{\nu\mathbf{Q}}=E^\text{bind}_\nu+\hbar^2(\mathbf{Q}-[\mathbf{K_c}-\mathbf{K_v}])^2/(2[m_c+m_v])+\varepsilon_{c\mathbf{K_c}}-\varepsilon_{v\mathbf{K_v}}$. Note, that exciton wavefunctions with different valley configurations are centered at different momenta. All necessary electronic band parameters can be found in \cite{kormanyos2015k}.

\begin{figure}[t!]
\includegraphics[width=80mm]{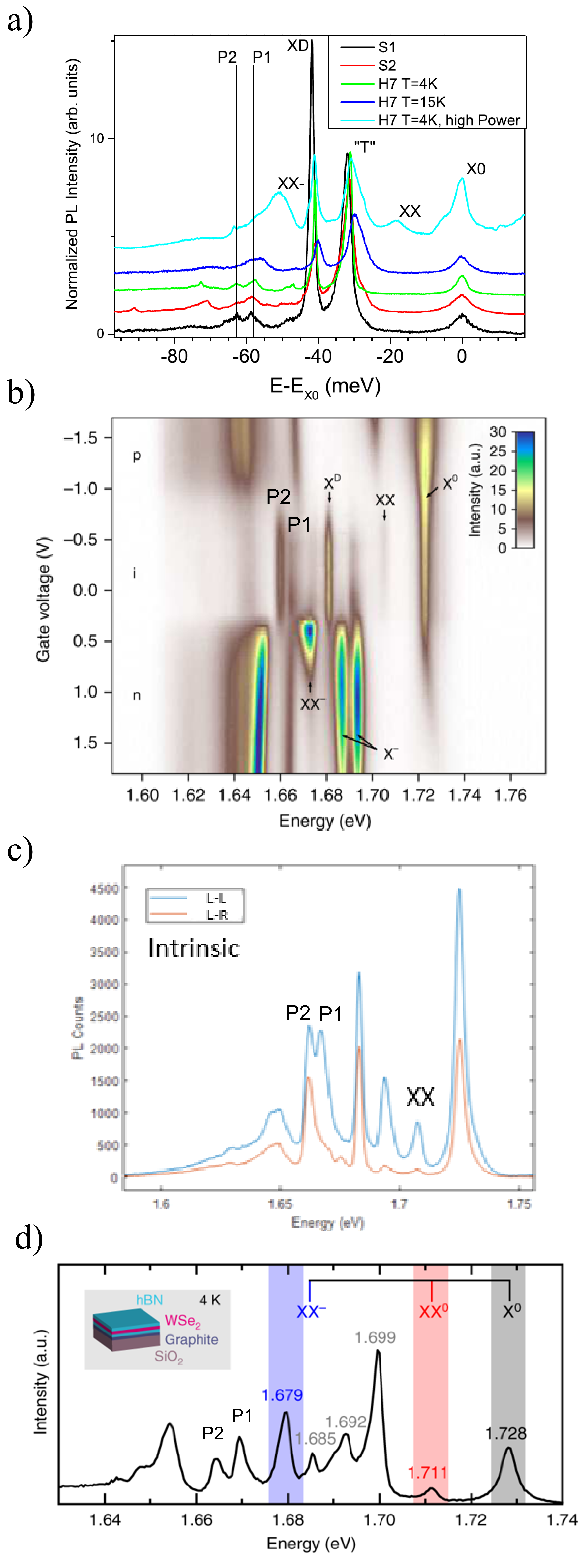}
\caption{Summary of three independent PL measurements on hBN-encapsulated WSe$_2$ monolayers at cryogenic temperatures. a) PL spectra from three different samples S1 (black) S2 (red) and H7 (green, blue, cyan). The following transitions are marked : bright neutral exciton X0, neutral and charged biexciton (XX,XX-), the so called trion (T), dark exciton (XD) and the transitions discussed in this work P1 and P2 b) Gate voltage dependence of the PL spectrum taken from ref. \onlinecite{ye2018efficient}. c) Polarization-resolved PL spectrum at charge neutrality point. L-L (L-R)
denotes the co-(cross)-polarized excitation and detection \cite{ye2018efficient}.
d) PL spectrum taken from ref. \onlinecite{barbone2018charge}. All three groups observed similar features in the PL about 60 meV below the bright exciton denoted with P1 and P2. Figure b), c) and d) are published unter the Creative Commons Attribution 4.0 International License, cf. https://creativecommons.org/licenses/by/4.0/. The labels P1 and P2 have been added.}
\label{fig:expComp} 
\end{figure}
\section{Experimental Observations}
Recent experimental PL studies on hBN-encapsulated tungsten diselenide have revealed a multitude of low-temperature emission peaks whose microscopic origin has not been fully clarified yet \cite{ courtade2017charged,lindlau2017identifying, ye2018efficient,barbone2018charge}. Most studies have so far focused on the impact of bound exciton configurations, such as trions, biexcitons and trapped excitons, while the potential influence of indirect phonon-assisted recombination of intrinsically dark exciton states has been ignored to a large extend. In Fig. \ref{fig:expComp} we have summarized the experimental observations of three independent measurements of the cryogenic luminescence from hBN-encapsulated WSe$_2$.

Fig. \ref{fig:expComp}a) shows PL spectra of different samples at charge neutrality. All three measured samples denoted with S1, S2 and H7 (same sample as used in main text) consistently exhibit several peaks below the bright exciton. In particular, the theoretically predicted signal denoted with P in the main text, which we attribute to the acoustic phonon assisted recombination of K-K' excitons, is systematically observed in form of two peaks denoted with P1 and P2 in Fig. \ref{fig:expComp}a)-c). The exact shape and intensity of the peaks P1 and P2 depends on excitation conditions (laser energy and power density) and temperature and we observe slight variations even for the same sample during different experiments. 
The same qualitative behaviour for the peaks P1 and P2 is observed in the measurements on very similar samples shown in Fig. \ref{fig:expComp}b) and d) which have been performed by other groups namely
Ye et.al. \cite{ye2018efficient} and Barbone et.al. \cite{barbone2018charge}, respectively. Note, that Fig. \ref{fig:expComp}b) presents a continuous study of the impact of the applied gate voltage, whereas d) is measured without gate voltage. Although the shape and relative intensity of the two peaks P1 and P2 seems to vary on different samples, they are clearly visible in all three measurements at a consistent energetic position, which agrees with the theoretically predicted peak stemming from acoustic phonon-assisted recombination of K-K' excitons, cf. main text. 

However, the signal in all three measurements appears as two separate peaks with a splitting of about 5 meV. Furthermore, it is striking that the energetic distance between P1 and P2 to the peak denoted as ``T'' in Fig. \ref{fig:expComp}a) corresponds approximately to the energy of KTO and KLO phonon, respectively. This suggests that these peaks could be phonon side bands of the ``T'' peak whose origin is still under debate. However, Fig. \ref{fig:expComp}c)  shows a polarization-resolved PL spectrum at the charge neutrality point measured by Ye et.al. (cf. supplementary of ref.\onlinecite{ye2018efficient}). Here a clear polarization dependence of those two peaks is visble, i.e. in cross-polarized case, only a single peak is observed. This polarization dependence indicates that the two peaks do not stem from the same exciton state, but might rather be a result of e.g. Coulomb exchange-mediated splitting of the two spin configurations of momentum-dark excitons. In particular, a mixing of momentum-dark states and direct K-K excitons could give rise to a Coulomb-exchange induced splitting of the K-K' and K'-K exciton, similar as in the case of spin-forbidden dark states \cite{robert2017fine}. In summary, further research is needed to understand the origin of the ``T'' peak and the possible impact of Coulomb exchange coupling on the splitting and hybridization of bright and dark exciton states.

\end{document}